\documentclass[%
twocolumn,
floatfix,
amsmath,amssymb,
aps,
superscriptaddress,
]{revtex4-2}

\usepackage[svgnames,psnames]{xcolor}
\usepackage{url}
\usepackage{graphicx}
\graphicspath{{.}{figs/}{../figs/}}
\usepackage{afterpage}
\usepackage{dcolumn}
\usepackage{bm}
\usepackage{tikz}
\usetikzlibrary{positioning,intersections,angles,quotes}
\usepackage[caption=false]{subfig}
\usepackage{booktabs} 
\usepackage{comment}
\usepackage{physics}
\usepackage{mathtools}

\usepackage[colorlinks,citecolor=DarkGreen,linkcolor=FireBrick,linktocpage,unicode,hypertexnames=false]{hyperref}
\usepackage[capitalize]{cleveref}
\crefname{enumi}{}{}
\usepackage{autonum}

\begin{document}

\title{
Sampling-based quasiprobability simulation for 
fault-tolerant quantum error correction 
on the surface codes under coherent noise
}

\author{Shigeo Hakkaku}
\email{shigeo.hakkaku@qc.ee.es.osaka-u.ac.jp}
\affiliation{%
  Graduate School of Engineering Science, Osaka University, 1-3 Machikaneyama, Toyonaka, Osaka 560-8531, Japan
}%
\author{Kosuke Mitarai}%
\email{mitarai@qc.ee.es.osaka-u.ac.jp}
\affiliation{%
  Graduate School of Engineering Science, Osaka University, 1-3 Machikaneyama, Toyonaka, Osaka 560-8531, Japan
}%
\affiliation{%
  Center for Quantum Information and Quantum Biology,
  Osaka University, 1-2 Machikaneyama, Toyonaka 560-0043, Japan
}%
\affiliation{%
  JST,
  PRESTO,
  4-1-8 Honcho, Kawaguchi, Saitama 332-0012, Japan
}%
\author{Keisuke Fujii}%
\email{fujii@qc.ee.es.osaka-u.ac.jp}
\affiliation{%
  Graduate School of Engineering Science, Osaka University, 1-3 Machikaneyama, Toyonaka, Osaka 560-8531, Japan
}%
\affiliation{%
  Center for Quantum Information and Quantum Biology,
  Osaka University, 1-2 Machikaneyama, Toyonaka 560-0043, Japan
}%
\affiliation{%
  RIKEN Center for Quantum Computing (RQC),
  Hirosawa 2-1, Wako, Saitama 351-0198, Japan
}%
\affiliation{%
  Fujitsu Quantum Computing Joint Research Division at QIQB,
  Osaka University, 1-2 Machikaneyama, Toyonaka 560-0043, Japan
}%
\date{\today}

\begin{abstract}
  We propose a sampling-based simulation for fault-tolerant quantum error correction under coherent noise. A mixture of incoherent and coherent noise, possibly due to over-rotation, is decomposed into Clifford channels with a quasiprobability distribution. 
  Then, an unbiased estimator of the logical error probability is constructed by sampling Clifford channels with an appropriate postprocessing. 
  We characterize the sampling cost via the channel robustness and find that the proposed sampling-based method is feasible even for planar surface codes with relatively large code distances intractable for full state-vector simulations.
  As a demonstration, we simulate repetitive faulty syndrome measurements on the planar surface code of distance 5 with 81 qubits. We find that the coherent error increases the logical error rate.
  This is a practical application of the quasiprobability simulation for a meaningful task and would be useful to explore experimental quantum error correction on the near-term quantum devices. 
\end{abstract}

\maketitle

\section{Introduction}
Quantum error correction (QEC) is an essential ingredient for developing scalable fault-tolerant quantum computers because quantum information is vulnerable to environmental noise \cite{shorSchemeReducingDecoherence1995,steaneMultipleparticleInterferenceQuantum1996}.
QEC counteracts noise by encoding quantum information into a subspace of multiple qubits, which assures computation with arbitrary precision in quantum computers.
Massive experimental efforts have been devoted to demonstrating small-scale QEC circuits as testbeds toward large-scale QEC circuits in the future as well as numerical simulations \cite{kellyStatePreservationRepetitive2015,eganFaulttolerantControlErrorcorrected2021,chenExponentialSuppressionBit2021}.
It is thus important to investigate performances of QEC codes theoretically to establish a plausible goal for experiments.

Most numerical studies for QEC have been conducted by 
assuming stochastic Pauli noise to exploit the efficient simulatability of stabilizer states \cite{gottesmanStabilizerCodesQuantum1997,aaronsonImprovedSimulationStabilizer2004}.
Specifically, Ref.~\cite{fowlerPracticalClassicalProcessing2012} numerically calculated the threshold error rate of the rotated surface code under single- and two-qubit depolarizing channels with circuit-level noise and observed a threshold error rate of 0.57\%.
This result suggests that the surface code can cope with the error rate that current state-of-the-art quantum computers are reaching \cite{jurcevicDemonstrationQuantumVolume2021,chenExponentialSuppressionBit2021}.
While the computational overhead increases when compared with the Pauli noise, we can also efficiently simulate the Clifford noise such as stochastic Clifford gates and Pauli projections.

In practice, however, quantum devices often suffer from noise that cannot be described by Clifford operations.
A major type of such noise is coherent unitary noise which is caused by the miscalibration of quantum gates which leads to over- or under-rotations.
Reference \cite{sheldonCharacterizingErrorsQubit2016} has developed a method to detect over-rotation errors using randomized benchmarking and detected $\pi/128$ over- or under-rotation errors in their superconducting qubit.
While the error has been calibrated subsequently in Ref.~\cite{sheldonCharacterizingErrorsQubit2016}, one can expect that a small amount of such errors beyond the experimental sensitivity are still present.

Analysis of the performance of QEC in such realistic situations still remains a challenge.
QEC circuits under non-Clifford noise have been investigated either by brute-force simulations \cite{tomitaLowdistanceSurfaceCodes2014, darmawanTensorNetworkSimulationsSurface2017} or by exploiting exact solvability of free fermion dynamics \cite{suzukiEfficientSimulationQuantum2017, bravyiCorrectingCoherentErrors2018}.
However, full state-vector simulations require exponential computational resources with respect to the code distance and are currently limited to distance-3 surface code which uses $25$ qubits \cite{tomitaLowdistanceSurfaceCodes2014}.
While the use of approximate simulation using a tensor network \cite{darmawanTensorNetworkSimulationsSurface2017} has pushed the limit to 153 qubits with perfect syndrome measurements, it is still difficult to scale up the simulation.
On the other hand, free fermion simulations can handle coherent errors in a scalable manner.
However, their usage is limited to certain cases: one-dimensional repetition codes
with faulty syndrome measurements~\cite{suzukiEfficientSimulationQuantum2017} which can only correct $X$ errors and surface codes with perfect syndrome measurements \cite{bravyiCorrectingCoherentErrors2018}.

In this paper, we propose a sampling-based simulation method widely applicable for fault-tolerant QEC circuits under a mixture of coherent and incoherent noise with multiple rounds of faulty syndrome measurements.
The central idea is to decompose (possibly non-Clifford) noise channels into the sum of completely stabilizer preserving (CSP) channels \cite{seddonQuantifyingMagicMultiqubit2019}. 
We simulate the circuits by sampling CSP channels according to quasiprobability distributions, which are obtained from the decompositions \cite{seddonQuantifyingMagicMultiqubit2019,benninkUnbiasedSimulationNearClifford2017}.
Each realization is efficiently simulable since the simulation of CSP channels involves only stabilizer states.
Note that Bennink \textit{et al.} have conducted similar simulations for small systems such as Steane's seven-qubit code~\cite{benninkUnbiasedSimulationNearClifford2017}.
We significantly improve the computational cost required for the simulation by providing more efficient decomposition of noise channels than Ref. \cite{benninkUnbiasedSimulationNearClifford2017}.
This reveals that we can perform an efficient simulation in the presence of coherent errors without any additional overhead for a wide range of practically interesting parameter regions.
Furthermore, even outside this region, the proposed quasiprobability method enables us to simulate 
a surface code of distance 5 with 81 qubits
on a single workstation within a reasonable computational time.

As demonstrations, we simulate the planar surface code under the code capacity coherent noise with distance up to $d= 7$ and under the phenomenological coherent noise with distance up to $d = 5$.
The result shows that such non-Clifford noise deteriorates the logical error rate as expected.
We also evaluated how many samples are required to simulate the logical error rate
reliably as a function of the noise parameters and the code distance.
This reveals that the proposed method allows us to simulate 
the planar surface code with relatively large code distances,
which are intractable for full state-vector simulations,
with a reasonable computational overhead.
The proposed method provides a benchmark for building small-scale fault-tolerant quantum computers in the noisy intermediate-sclae quantum (NISQ) era.

\section{Simulation of QEC circuits under coherent noise}\label{sec:alg}
In this section, we discuss how to calculate a logical error rate of a QEC code by simulating quantum circuits with a quasiprobability sampling of CSP channels.
QEC requires two types of qubits: data qubits, which constitute logical qubits, and measurement qubits, which are used for detecting errors on data qubits.
The measurements extract eigenvalues of code stabilizers by measuring the latter, and these eigenvalues are called error syndromes.
For a distance $d$ code, we repeat such measurements for $d$ rounds. 
We use a (noisy) Clifford circuit $\mathcal{E}_\text{synd}$ for the repetitive syndrome measurements.
The measurement qubits of different rounds are to be treated as different qubits to simplify the notation.
Let $b$ be the error syndrome in space and time.
When data qubits are initialized to $\ket{0_L}$, the probability of obtaining a specific  error syndrome $b$ is given by
\begin{align}
    p(b) 
    &= \mel**{b}{\Tr_\text{data}\bqty{\mathcal{E}_\text{synd}\pqty{\op{\vb{0}}}}}{b}. \label{eq:prob_of_syndrome}
\end{align}
where $\ket{b}$ and $\ket*{0^{\abs{b}}}$ are final and initial states of the measurement qubits, respectively, and $\ket{\vb{0}}\coloneqq \ket{0_L}\otimes \ket{0^{\abs{b}}}$
After the extraction of the error syndrome, we feed $b$ to decoding algorithms such as a minimum-weight perfect-matching algorithm to find a possible recovery operation $\mathcal{R}_b$ which corrects errors on data qubits.
The error corrected state $\rho_\text{corrected}$ is given by
\begin{align}
    \rho_\text{corrected} = \sum_b \mathcal{R}_b \circ \mathcal{P}_b \circ \mathcal{E}_\text{synd}\pqty{\op{\vb{0}}} \label{eq:corrected_state},
\end{align}
where $\mathcal{P}_b$ is the projection onto $\op{b}$.
Since the date qubits of $\rho _{\rm corrected}$ are in the code space, the logical fidelity can be expressed by $\mel{0_L}{\Tr_\text{meas}\pqty{\rho_\text{corrected}}}{0_L}$.
Therefore the logical error rate $p_L$ can be written as
\begin{align}
    p_L &= 1 - \mel{0_L}{\Tr_\text{meas}\pqty{\rho_\text{corrected}}}{0_L} \\
    &= 1 - \sum_b \mel{0_L}{\Tr_\text{meas}\pqty{\mathcal{R}_b \circ \mathcal{P}_b \circ \mathcal{E}_\text{synd}\pqty{\op{\vb{0}}}}}{0_L} \label{eq:pl}.
\end{align}
If the noise introduced in $\mathcal{E}_\text{synd}$ is a stochastic Pauli or Clifford error, one can simulate $\mathcal{E}_\text{synd}$ efficiently and can estimate the logical error rate $p_L$.
However, efficient simulatability vanishes if noise involves non-Clifford channels.

We now describe an idea to deal with more general noise by a quasiprobability method \cite{stahlkeQuantumInterferenceResource2014,pashayanEstimatingOutcomeProbabilities2015,howardApplicationResourceTheory2017,benninkUnbiasedSimulationNearClifford2017,seddonQuantifyingMagicMultiqubit2019,hakkakuComparativeStudySamplingBased2021,mitaraiOverheadSimulatingNonlocal2021,seddonQuantifyingQuantumSpeedups2021}.
$\mathcal{E}_\text{synd}$ can be decomposed into (noisy) elementary operations as $\mathcal{E}_\text{synd} = \mathcal{E}^{(L)} \circ \cdots \circ \mathcal{E}^{(1)}$.
Here $L$ is the total number of quantum operations in $\mathcal{E}_\text{synd}$.
$\mathcal{E}^{(i)}$ can be decomposed over CSP and completely positive trace-preserving (CPTP) channels $\mathcal{S}_k^{(i)}$ in terms of a quasiprobability distribution $c^{(i)}_{k}$ as
\begin{align}
    \mathcal{E}^{(i)} &= \sum_{k} c^{(i)}_{k} \mathcal{S}_{k}^{(i)}.
\end{align}
This decomposition can alternatively be written as
\begin{align}
                  \mathcal{E}^{(i)}
                  &= \sum_{k} p_{k}^{(i)} R_*\pqty{\mathcal{E}^{(i)}} \text{sgn}\pqty{c_{k}^{(i)}}  \mathcal{S}_{k}^{(i)}, \label{eq:deco_non_ch_robustness}
\end{align}
where
\begin{align}
    R_*\pqty{\mathcal{E}} &\coloneqq \min_{\Bqty{c_k}}\Bqty{\sum_k\abs{c_k};\mathcal{E} = \sum_k c_k \mathcal{S}_k},\\
    p_k^{(i)} &\coloneqq \frac{\abs{c_k^{(i)}}}{R_*\pqty{\mathcal{E}^{(i)}}}.
\end{align}
$R_*\pqty{\mathcal{E}}$ is called channel robustness \cite{seddonQuantifyingMagicMultiqubit2019}, the square of which characterizes the sampling cost, as will be seen later.
Using this decomposition for each $\mathcal{E}^{(i)}$,  $\mathcal{E}_{\text{synd}}$ becomes
\begin{align}
\mathcal{E}_\text{synd}
&= \sum_{\vec{k}} p_{\vec{k}} R_{*\text{tot}}\lambda_{\vec{k}}\mathcal{S}_{\vec{k}},\label{eq:E_synd_decomposition}
\end{align}
where the summation is taken over all possible $\vec{k} = \pqty{k^{(1)}, k^{(2)}, \ldots , k^{(L)}}$, and
\begin{align}
    p_{\vec{k}} &\coloneqq \prod_{i=1}^L p_{k^{(i)}}^{(i)},\\
    \lambda_{\vec{k}} &\coloneqq \prod_{i=1}^L \text{sgn} \pqty{c_{k^{(i)}}^{(i)}}, \\
    \mathcal{S}_{\vec{k}} &\coloneqq \mathcal{S}^{(L)}_{k^{(L)}} \circ \cdots \circ \mathcal{S}_{k^{(1)}}^{(1)}, \\
    R_{*\text{tot}} &\coloneqq \prod_{i=1}^L R_{*}\pqty{\mathcal{E}^{(i)}}.
\end{align}
Finally, combining \cref{eq:pl,eq:E_synd_decomposition}, we conclude
\begin{align}
    p_L &= 1 - \sum_{b,\vec{k}} p_{\vec{k}} R_{*\text{tot}} \lambda_{\vec{k}} \mel{0_L}{\Tr_\text{meas}\pqty{\mathcal{R}_b\circ \mathcal{P}_b \circ \mathcal{S}_{\vec{k}}\pqty{\op{\vb{0}}}}}{0_L}. \label{eq:pl_weak}
\end{align}
This implies that, when $\vec{k}$ is sampled from $p_{\vec{k}}$, $1-R_{*\text{tot}}\lambda_{\vec{k}}\mel{0_L}{\Tr_\text{meas}\pqty{\mathcal{R}_b\circ \mathcal{P}_b \circ \mathcal{S}_{\vec{k}}\pqty{\op{\vb{0}}}}}{0_L}$ is an unbiased estimator for $p_L$.
Since it is bounded in a range $[-R_{*\text{tot}}, R_{*\text{tot}}]$, 
from Hoeffding inequality~\cite{hoeffdingProbabilityInequalitiesSums1963}, the number of samples $M$ needed to estimate $p_L$ within additive error $\epsilon$ with probability at least $1-\delta$ is given by
\begin{align}
    M = \frac{2}{\epsilon^2}R_{*\text{tot}}^2 \ln\pqty{\frac{2}{\delta}}. \label{eq:num_samples}
\end{align}
Note that, when we only consider Clifford noise, $\frac{2}{\epsilon^2}\ln\pqty{\frac{2}{\delta}}$ samples suffice to achieve the same accuracy.
Therefore $R_{*\text{tot}}^2$  quantifies the additional overhead required for including the effect of non-Clifford channels.

\section{Planar surface codes under coherent noise} \label{sec:surface_codes_under_rotation}
\begin{figure}[tb]
    \includegraphics[width=.65\linewidth]{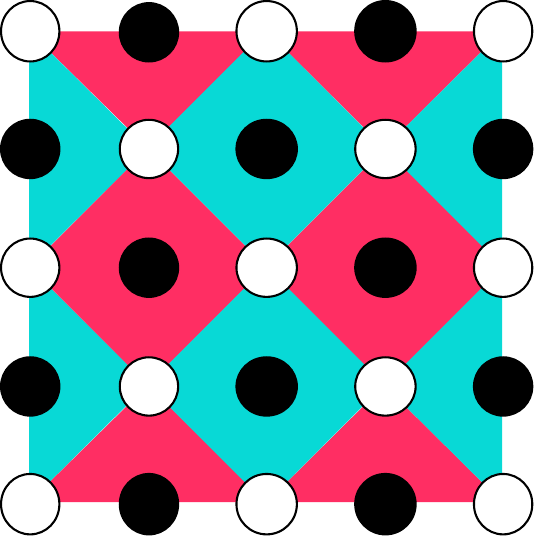}
    \caption{Layout of $d=3$ planar surface code.
    The white and black circles represent data and measurement qubits, respectively.
    The light blue square and triangular patches show $X$ stabilizers.
    The red square and triangular patches show $Z$ stabilizers.}
    \label{fig:surface_code}
\end{figure}
To demonstrate the feasibility of the proposed method, 
we consider the planar surface code introduced in Ref.~\cite{dennisTopologicalQuantumMemory2002}, 
which is thought to be one of the most promising candidates for an experimental realization of QEC, as they require only single- and nearest-neighbor two-qubit gates on two-dimensional arrays of qubits \cite{kitaevFaulttolerantQuantumComputation2003,fowlerPracticalClassicalProcessing2012}.
The planar surface code with code distance $d$ has a $(2d-1)\times (2d-1)$ square grid of qubits of which $d^2 + (d-1)^2$ data qubits are used to encode the logical qubit and $2d(d-1)$ measurement qubits are used to extract the syndromes.
In \cref{fig:surface_code}, we show the layout of $d=3$ planar surface code as an example.

In numerical simulations, the ideal logical state $\ket{0}_L$ followed by single-qubit noise is prepared as the initial state.
We assume two types of noise model: 
a code-capacity noise model, where the noise occurs in all data qubits with perfect syndrome measurements, and a phenomenological noise model, where the noise occurs in all data qubits and measurement qubits just before the syndrome measurements.
The number of rounds of the syndrome measurement in the latter case is $d$.
We also assume that the syndrome measurements are performed perfectly at the final cycle.
In both cases, $Z$-type and $X$-type errors are uncorrelated, and hence 
only $X$-type errors and syndrome measurements are simulated for simplicity.
The specific noise channel $\mathcal{N}_\text{coh}$ simulated in this paper is a mixture of coherent and incoherent noise which is modeled by the over-rotation noise followed by the bit-flip $X$ error as
\begin{align}
    \mathcal{N}_\text{coh} &\coloneqq \mathcal{N}_\text{bit-flip} \circ \mathcal{N}_\text{over-rot}, \\
    \mathcal{N}_\text{over-rot} &\coloneqq \bqty{e^{ir\theta X}}, \\
    \mathcal{N}_\text{bit-flip} &\coloneqq \pqty{1-p} \bqty{I} + p \bqty{X},
\end{align}
where $\theta$ is chosen such that $p = \sin^2\theta$.
We vary the parameters $(r, p)$ and evaluate the performance of the code by using the method described in Sec.~\ref{sec:alg}.

\begin{figure}[tb]
    \includegraphics[width=\linewidth]{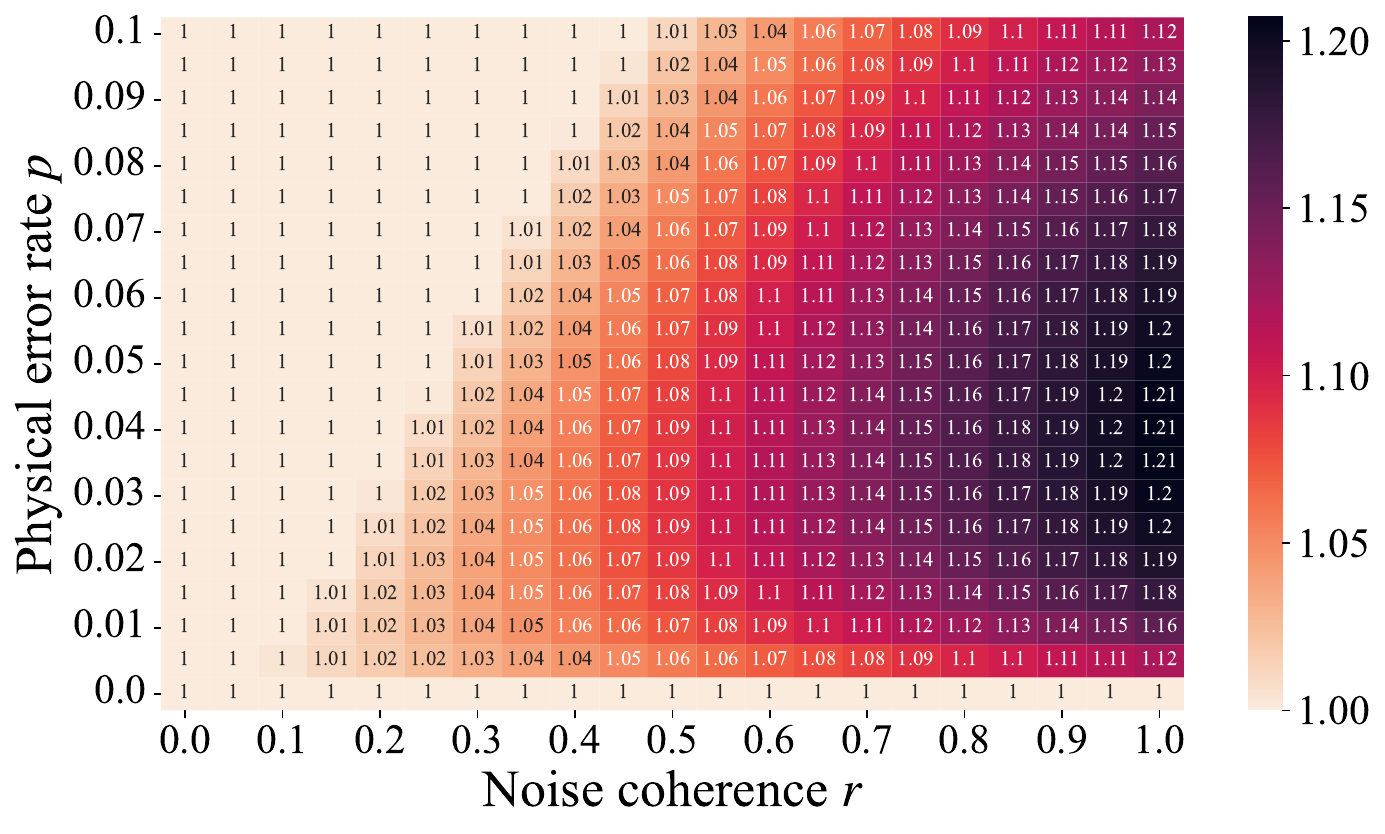}
    \caption{Channel robustness of the coherent noise $R_\text{*coh}\pqty{r,p}$.
        The horizontal and vertical axes display the noise coherence $r$ and physical error rate $p$, respectively.
        Numbers in each cell show the value of $R_\text{*coh}\pqty{r,p}$.
        }
    \label{fig:ch_robustness_heatmap}
\end{figure}

We first examine the sampling cost of our simulation which is characterized by the channel robustness $R_\text{*coh}\pqty{r,p}$ of $\mathcal{N}_\text{coh}$~\cite{howardApplicationResourceTheory2017,hakkakuComparativeStudySamplingBased2021,seddonQuantifyingMagicMultiqubit2019}.
The CSP channels employed to decompose $\mathcal{N}_\text{coh}$ are $\bqty{I}$, $\bqty{X}$, $\bqty{e^{-i (\pi /4)X}}$, and $\bqty{Xe^{-i (\pi /4)X}}$.
\Cref{fig:ch_robustness_heatmap} shows the values of  $R_\text{*coh}\pqty{r,p}$.
From \cref{fig:ch_robustness_heatmap}, we confirm that the channel robustness increases as the noise coherence becomes larger as expected.
Importantly, for a small $r$ with a sufficiently large $p$, the channel robustness decreases and hits unity, where an efficient simulation of coherent errors can be performed.
This is because
$r\theta$ is small in this region, resulting in the low channel robustness of $\mathcal{N}_\text{over-rot}$.
Thus, the bit-flip noise with large $p$ can easily make the channel robustness unity.
On the other hand, at $p=0$, the channel robustness of $\mathcal{N}_\text{over-rot}$ is unity even if $r>0$ since we set $p=\sin^2\theta$.
Therefore a certain mixture of incoherent and coherent errors, for example,
with $p=1\%$ ($0.1\%$) 
and $r = 0.10$ ($r=0.46$),
can be efficiently simulated without any additional overhead,
which is in an experimentally important parameter region.
This greatly improves the simulation cost over Ref. \cite{benninkUnbiasedSimulationNearClifford2017} which is a result of decomposing $\mathcal{N}_\text{coh}$ as a whole rather than decomposing $\mathcal{N}_\text{over-rot}$ and $\mathcal{N}_\text{bit-flip}$ individually.

The number of samples needed for accurate results is determined by $R_{*\text{tot}}^2$ via \cref{eq:num_samples}.
In the case of the code capacity noise, $\mathcal{N}_{\text{coh}}$ is applied $d^2+(d-1)^2$ times, which corresponds to the number of data qubits.
Therefore $R_{*\text{tot}}^2=\pqty{R_\text{*coh}\pqty{r,p}}^{2\pqty{d^2+(d-1)^2}}$.
In the case of the phenomenological noise, $\mathcal{N}_{\text{coh}}$ is applied to each of the $d^2+(d-1)^2$ data qubits $d$ times and to each of the $d\pqty{d-1}^2$ measurement qubits for $X$-type errors $d-1$ times since we assume perfect measurement in the final round.
Overall, $\mathcal{N}_{\text{coh}}$ is applied $d\pqty{3d^2-4d+2}$ times, which means $R_{*\text{tot}}^2 = \pqty{R_\text{*coh}\pqty{r,p}}^{2d\pqty{3d^2-4d+2}}$ in this case.
These formulas for $R_{*\text{tot}}^2$ provide us estimates of the simulation cost for a given $(p,r,d)$, based on which we choose the parameter range investigated below.

\begin{figure}[tb]
\subfloat[\label{subfig:pl_code_capacity}]{%
  \includegraphics[width=\linewidth]{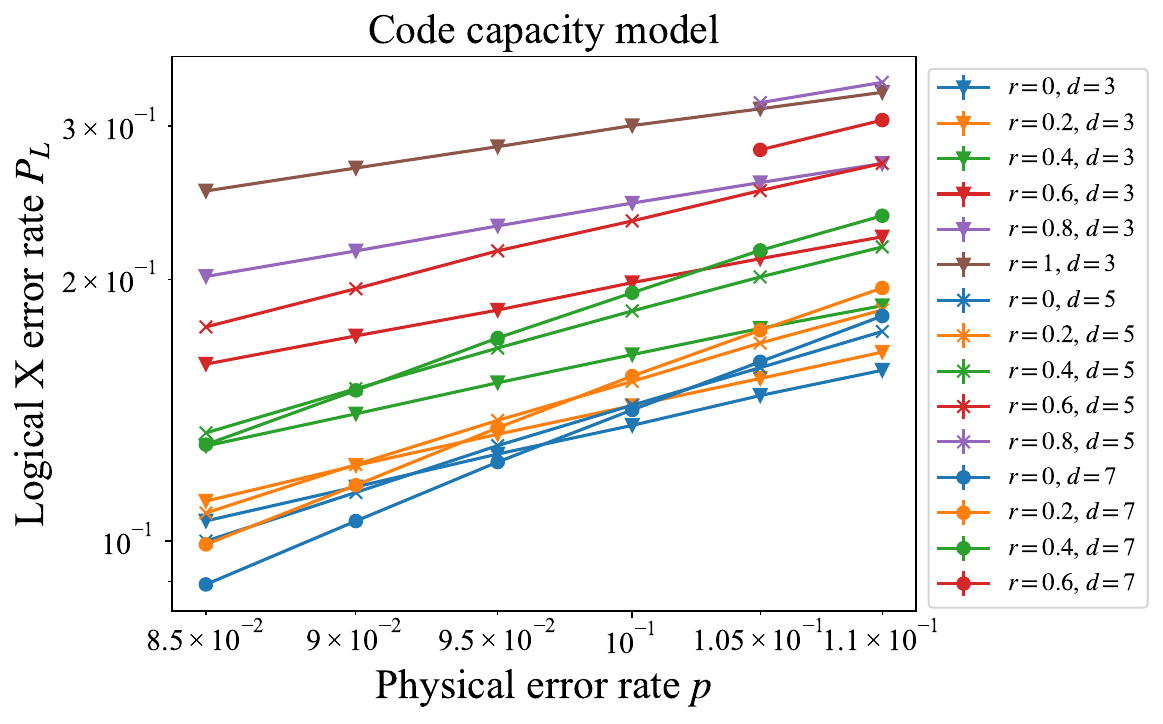}%
}\hfill
\subfloat[\label{subfig:pl_phenomenological}]{%
  \includegraphics[width=\linewidth]{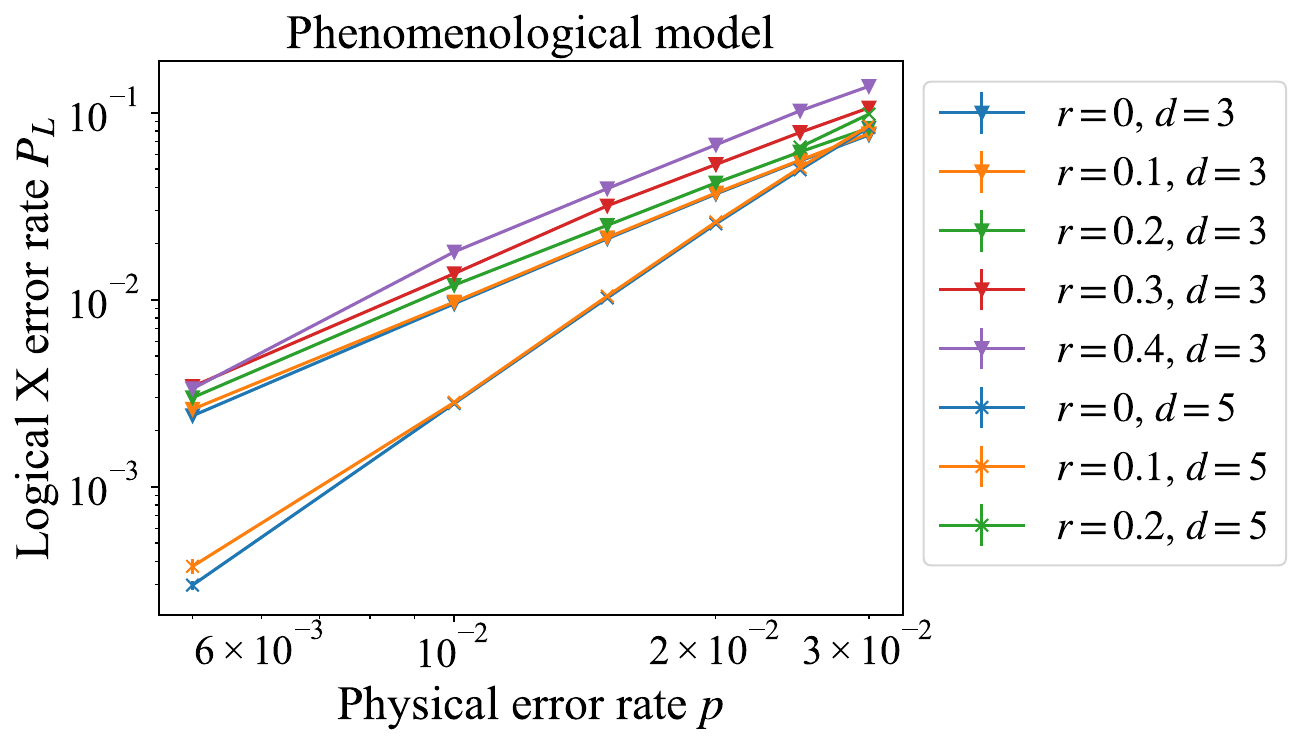}%
}
\caption{Logical $X$ error rate of the planar surface code under the coherent noise as a function of the physical error rate $p$ and noise coherence $r$ in the case of code capacity \protect\subref{subfig:pl_code_capacity} and phenomenological \protect\subref{subfig:pl_phenomenological} noise.
The horizontal axis shows the physical error rate $p$, and the vertical axis shows the logical $X$ error rate $p_L$.
The triangles, crosses, and circles stand for $d=3, 5, 7$, respectively.
The color shows the noise coherence.
}
\label{fig:pl}
\end{figure}
\begin{figure}[tb]
    \includegraphics[width=\linewidth]{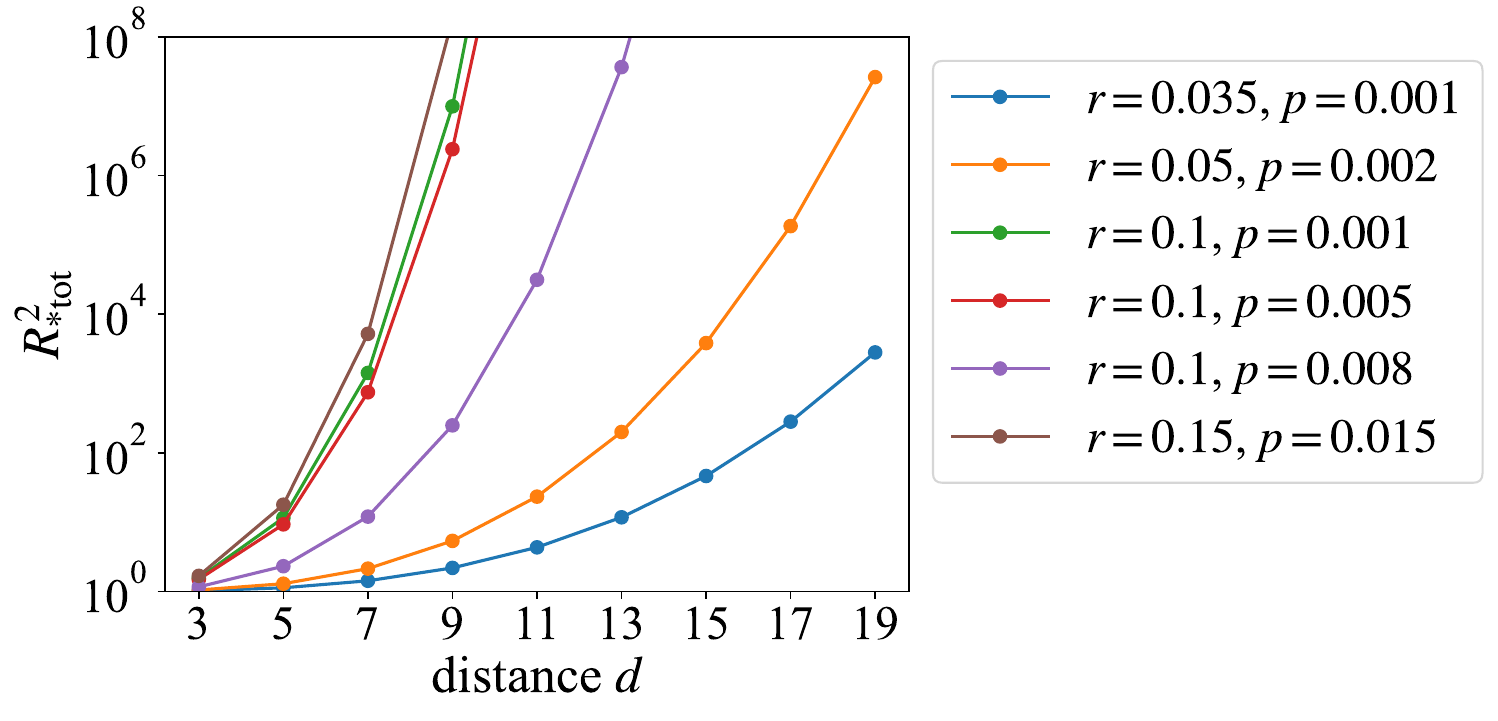}
    \caption{Scaling of the overhead caused by the quasiprobability sampling, $R^2_{\text{*tot}}$, as a function of the code distance $d$.
    }
    \label{fig:scaling_of_ch_robustness}
\end{figure}

\Cref{fig:pl} shows the logical error rate $p_L$ as a function of physical error rate $p$ and noise coherence $r$, where the parameters are chosen such that our workstation with Intel Xeon CPU v4 CPU (E5-2687W), 24 cores, 3.00GHz, can calculate each point within a few days at most.
We confirmed that the standard error of each data point is below $10^{-3}$.
From \cref{fig:pl}, the logical error rate increases as the noise coherence grows, which implies that the impact of the coherent noise on the logical error probability is not negligible even for a relatively large code distance.
Note that the $d=5$ code, which requires 81 qubits, is well beyond the reach of naive full state-vector simulation.
Furthermore, it is the first analysis of this region with faulty syndrome measurements to the best of our knowledge.
Finally, let us discuss with which parameters and code distance the 
proposed method works.
\Cref{fig:scaling_of_ch_robustness} shows the dependence of $R_\text{*tot}^2$ with respect to $d$ for the phenomenological noise model.
Note that for the parameters where $R_\text{*coh}\pqty{r,p}=1$ in \cref{fig:ch_robustness_heatmap}, we can simulate without any additional overhead as mentioned before. 
We will be able to simulate large code distances in that region.
Outside of that, we expect that regions with $R_\text{*tot}<10^3$ are within reach if a high-performance parallel computer of $10^6$ CPU cores is available.
For example, realistic parameters such as $(p,r,d)=(1.5\%,0.15,7)$ and $=(0.2\%,0.05,13)$ result in $R_\text{*tot}<10^3$.
Full state-vector simulation would not work for these numbers of qubits; we need
169 qubits for $d=7$ and 625 qubits for $d=13$.

\section{Conclusion} \label{sec:conclusion}
We have proposed a sampling-based method to estimate the logical error rate of QEC codes under coherent noise such as an over-rotation error.
The simulation protocol is based on the quasiprobability decomposition of  noise channels into Clifford operations.
It is interesting to note that 
the QEC process is simulated as usual 
for sampled CSP channels, and hence 
the probability distribution for the syndrome measurements is far different from the true one $p(b)$.
However, if we sample whether the decoding
successds or fails with the quasiprobability method, we can estimate the logical error rate.
By calculating the channel robustness for the 
mixture of coherent and incoherent errors,
we reduce the simulation costs substantially, 
which allows us to simulate a practically important 
parameter region with a relatively large code distance 
without any additional overhead or with a reasonable additional overhead.
While we have only considered the phenomenological noise model,
it is straightforward to extend our method to the circuit-level noise model, where each elementary gate is followed by noise.
We leave these problems for future works.
We believe that this work helps to analyze the performance of the near-term small-scale QEC in realistic situations.

\begin{acknowledgements}
K.M. is supported by JST PRESTO Grant No. JPMJPR2019 and JSPS KAKENHI Grant No. 20K22330.
K.F. is supported by JST ERATO Grant No.~JPMJER1601 and JST CREST Grant No.~JPMJCR1673.
This work is supported by MEXT Quantum Leap Flagship Program (MEXT QLEAP) Grants No.~JPMXS0118067394 and No.~JPMXS0120319794.
This work was supported by JST Moonshot R\&D Grant No.~JPMJMS2061.
We also acknowledge support from the JST COI-NEXT program.
We thank Sho Takagi and Mitsuki Katsuda for valuable discussions about surface codes and the implementation of minimum-weight perfect matching (MWPM) decoder.
We also thank Yasunari Suzuki for the implementation of the Aaronson and Gottesman's CNOT-Hadamard-phase (CHP) simulator.
\end{acknowledgements}
\bibliography{sampling_simulation_of_surface_codes}
\end{document}